\documentclass[%
 reprint,
 amsmath,amssymb,
 aps,
floatfix,
]{revtex4-2}

\usepackage{graphicx}% Include figure files
\usepackage{dcolumn}% Align table columns on decimal point
\usepackage{bm}% bold math
\usepackage{float}
\usepackage[hidelinks]{hyperref}
\usepackage{color}
\usepackage{upgreek}
\usepackage{comment}
\usepackage{svg}

\newcommand{\um}{$\upmu$m } % Keep the space otherwise they stick

\newcommand{\amax}{\alpha_\text{max}}

\begin{document}

\preprint{APS/123-QED}

\title{Fabricating fiber cavity mirror substrates compatible with high coupling efficiency}

\author{Michael Caouette-Mansour}
\affiliation{%
Department of Physics, McGill University, 3600 Rue University, Montreal QC, H3A 2T8, Canada
}%
\author{Thomas J. Clark}
\affiliation{%
Department of Physics, McGill University, 3600 Rue University, Montreal QC, H3A 2T8, Canada
}%
\author{Valeria Mosso Tsedilkina}
\affiliation{%
Department of Physics, McGill University, 3600 Rue University, Montreal QC, H3A 2T8, Canada
}%
\author{Jack Sankey }
\affiliation{%
Department of Physics, McGill University, 3600 Rue University, Montreal QC, H3A 2T8, Canada
}%

\date{\today}% It is always \today, today,

\begin{abstract}

Fiber optical cavities offer small mode volumes and correspondingly strong light-matter interactions in an open Fabry-Perot geometry. However, existing fabrication techniques do not reliably produce substrates with surface profiles amenable to high mode matching between the cavity mode and fiber core, thereby limiting the achievable collection efficiency. Here we present a technique to fabricate fiber mirror substrates while using \textit{in situ} reflectometry to constrain the achievable mode matching prior to coating. By measuring the back-reflection from freshly cleaved fiber tips, we pre-select 138 fibers compatible with 96.5-99.5\% mode matching, and after a single CO$_2$ laser ablation pulse, these fibers remained compatible with 95.3-99.2\%. This simple technique provides rapid feedback during each stage of substrate fabrication, greatly enhancing the yield of viable fiber mirror substrates prior to (expensive) coating runs.

\end{abstract}
\maketitle

%\tableofcontents
\section{Introduction}
% Few words on Fiber FP cavity 
% And why is the mode coupling important.
Fiber Fabry-Perot optical cavities \cite{hunger2010a} are now a common tool for enhancing light-matter interactions \cite{pfeifer_achievements_2022}, advantageously offering mode volumes with micron dimensions and direct coupling between the cavity and fiber core modes. A dominant limitation in these systems is the achievable collection efficiency -- the fraction of cavity light that is coupled to the traveling mode of the fiber core. This is generally limited by losses due to absorption, scattering, ``clipping'' \cite{benedikter_transverse-mode_2015}, and (the focus of this work) coupling to the fiber \textit{cladding} arising from mismatch between the cavity and fiber core mode profiles at the mirror surface (Fig.~\ref{fig1:prms_geometry}(a)). In applications requiring high collection efficiency, especially those involving fragile quantum states (e.g., quantum networks \cite{volz_measurement_2011, monroe_large-scale_2014, brekenfeld_quantum_2020, niemietz_nondestructive_2021, awschalom_development_2021, covey_quantum_2023}, quantum optic devices \cite{wagner_direct_2018, gallego_strong_2018, fernandez-gonzalvo_fully_2023, zifkin_lifetime_2024, hansen_optical_2025} and squeezing \cite{brieussel_toward_2018, takanashi_generation_2019, takanashi_4-db_2020, mcgarry_microstructured_2024}), these loss channels introduce additional quantum noise \cite{clerk2010introduction}, thereby placing bounds on what is possible to achieve.

A particularly challenging limitation is the cavity-core mode matching parameter $\eta_T$ \cite{gallego_high-finesse_2016}(Fig.~\ref{fig1:prms_geometry}(a)). To maximally match the cavity and core mode profiles -- thereby achieving the highest possible $\eta_T$ -- the mirror surface should be flat, with its normal vector aligned to the fiber core. At the same time, the cavity mode (defined by the location and shape of the second mirror) should have its waist centered on the core at this surface, with a mode field diameter equal to that of the core mode. 

The simplest fabrication paradigm \cite{hunger2010a, garcia_dual-wavelength_2018, ruelle_optimized_2019, gao_profile_2025} is to first cleave the optical fiber, leaving a ``flat'' surface with $\sim$nm roughness (see below), and then ablate the surface with a milliseconds-long pulse of 10.6-\um-wavelength light from a CO$_2$ laser, leaving a dimpled surface near the core with roughness $\lesssim$0.2~nm \cite{hunger2010a}. This smooth substrate can then be coated with a low-loss dielectric Bragg mirror. 

Cleave errors and ablation errors both erode the achievable value of $\eta_T$. As discussed below, a high-quality cleaver can usually produce surfaces compatible with $\eta_T>0.95$, leaving the dominant challenge of simultaneously aligning the ablated dimple with the fiber core and achieving a sufficiently large radius of curvature (i.e., approaching the ``flat'' limit). When combined, these issues generally limit mode-matching to $\lesssim$85\% in practice \cite{hunger2010a}. 
A better control on the fiber tip geometry has been achieved with multi-shot ablation protocols \cite{ott_millimeter-long_2016, garcia_dual-wavelength_2018}, or one can increase the core's mode field diameter using photonic crystal fibers to match modes with larger waists \cite{ott_millimeter-long_2016}. However, improvement of mode matching above 90\% has only been achieved by introducing additional complexity, such as splicing an additional bulky optic to the fiber tip \cite{gulati_fiber_2017} and incorporating metamaterials and free-space optics \cite{wang_coupling_2025}, requiring additional steps to integrate with standard (e.g., telecom) fibers.

Here we exploit the stable, low-coherence light from a superluminescent diode to demonstrate a simple, high-precision reflectometry technique capable of constraining the quality of a fiber substrate surface during each stage of fabrication. In particular, this readily allows the pre-selection of substrates compatible with $\eta_T > 0.95$ -- using the simplest single-ablation protocol -- prior to the (expensive \& time-consuming) coating process. We first present a simple model describing the relationship between measured back-reflection and the range of possible mode matching parameters, then use this to characterize 138 fiber substrates, tracking their evolution through the cleave and ablation steps. By combining these reflectometry measurements with optical profilometry, we are able to further constrain the achievable $\eta_T$, finding a high yield of fiber substrates compatible with $\eta_T>0.95$. Notably, in this high-$\eta_T$ limit, we find that reflectometry alone places tight bounds (less than $\pm$0.1\%) on the achievable $\eta_T$, eliminating the need for a cumbersome profilometry step.

\section{Modeling reflection and mismatch}
\label{sec:theory}

\begin{figure}
\includegraphics[scale=0.9]{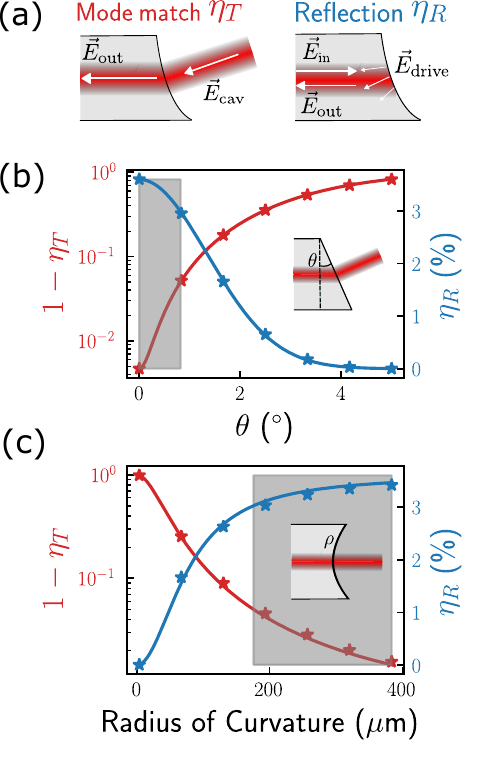}
\caption{
\label{fig1:prms_geometry}
\textbf{Coupling efficiency and reflection coefficient}
(a) Diagrams showing the geometry for (left) an optimal cavity mode (field $\vec E_\text{cav}$, red intensity profile normal to the surface) transmitting into a fiber core mode ($\vec E_\text{out}$) with mode matching parameter $\eta_T$, and (right) an input core mode ($\vec E_\text{in}$) reflecting back into the core mode ($\vec E_\text{out}$) via the induced ``drive'' field $\vec E_\text{drive}$ at the surface.
(b,c) Relationship between the mode \textit{mismatch} parameter $1-\eta_T$ (red) and the reflection \textit{matching} parameter $\eta_R$ (blue) as a function of (b) cleave angle $\theta$ for the simple case of a flat surface (see inset) and (c) radius of curvature $\rho$ for the simple case of the dimple perfectly centered (see inset). The red (blue) markers are the result of numerical integration of Eq.~\ref{eq:epsilon_def} (Eq.~\ref{eq:R_def}) with the Bessel solutions for the single-mode core \cite{saleh_fundamentals_2019}. The shaded region corresponds to surfaces compatible with $\eta_T>0.95$. All shown curves are monotonic, motivating a one-to-one mapping between measured reflection $\eta_R$ and $\eta_T$. As discussed below, when allowing both $\theta$ and $\rho$ to vary, $\eta_R$ still strongly constrains $\eta_T$ in the desired high-$\eta_T$ limit.
}
\end{figure}

In a fiber cavity, the field profile of the cavity mode $\vec{E}_\text{cav}$ at the fiber surface generally does not match that of the outbound core mode $\vec E_\text{out}$ (see Fig.~\ref{fig1:prms_geometry}(a)). As a result, only a fraction
\begin{align}
\label{eq:epsilon_def}
    \eta_T &= \left|\langle \vec{E}_\text{cav}|\vec{E}_{\text{out}}\rangle \right|^{2}
    =\left| \sum_{j\in \{x,y,z\}}\langle E_{\text{cav},j}|E_{\text{out},j}\rangle\right|^2
\end{align}
of light transmitted through the mirror coating actually transfers to the fiber core. The second form explicitly writes out the dot products in terms of field components $j\in \{x,y,z\}$ for reference, where the fields are integrated along the surface, and all fields $\vec E$ are normalized so that the inner product $\langle \vec E|\vec E\rangle=1$.

The field in a typical single-mode fiber is well-approximated as a plane wave with nearly Gaussian intensity profile \cite{saleh_fundamentals_2019}, and so $\eta_T$ is maximized when the mirror surface is parallel to the traveling mode's (flat) phase fronts, and the cavity mode (assuming Gaussian intensity profile) has its waist centered on the core at the surface, with an optimally matched mode field diameter. For a common step-index telecom fiber (SMF-28 Ultra, e.g.) at 1550 nm, $\eta_T$ can be as high as $\eta_{T,\max} = 0.995$. Graded index fibers may improve upon this, but this limit is already far beyond what is needed for near-term applications ($\eta_T\gtrsim 0.95$) such as squeezing, since $1-\eta_T=5\%$ loss is already compatible with $\gtrsim 13$ dB squeezing \cite{bachor_guide_2004}.

As mentioned, the simplest method for fabricating smooth fiber mirror substrates is to cleave an optical fiber and ablate the resulting surface with a CO$_2$ laser pulse \cite{hunger2010a}. This melts the glass, allowing surface tension to enforce angstrom-scale roughness when it re-solidifies, and necessarily also evaporates some material, thereby altering the surface profile. Together with cleave imperfections and $\sim$$\upmu$m tolerance in the manufactured position of the fiber core relative to the cladding, this generates errors in the fabricated surface profile that ultimately limit the maximum achievable value of $\eta_T$. To leading order, the errors are the surface's ``tilt'' angle $\theta$ (Fig.~\ref{fig1:prms_geometry}(b) inset) and its radius of curvature $\rho$ (Fig.~\ref{fig1:prms_geometry}(c) inset). 

To get a sense of scale, Fig.~\ref{fig1:prms_geometry}(b) shows the mode \textit{mismatch} parameter $1-\eta_T$ (red curve) for a flat surface tilted by angle $\theta$ (see inset) assuming an optimally situated cavity mode. Approximating the fiber mode with a Gaussian profile of radius $\sigma$ \cite{joyce_alignment_1984} yields
\begin{equation}
\eta_T \approx \eta_{T,\max} e^{-\sigma^{2}\beta^{2}\theta^{2}/4}, 
\end{equation}
to leading order in $\theta$, where $\beta$ is the propagation constant (effective wavenumber) along the fiber core. Normalizing by $\eta_{T,\max}$ improves agreement with numerical integration using the true fiber core mode (red markers). Evidently, tilt errors below 0.8$^\circ$ (shaded region) can achieve $\eta_T>95$\%. Similarly, if there is no tilt but there is a radius of curvature $\rho$ (Fig.~\ref{fig1:prms_geometry}(c)),
\begin{equation}
\eta_{T} \approx \frac{\eta_{T,\text{max}}}{1+\sigma^{4}\beta^{2}/16 \rho^{2}},
\end{equation}
to leading order in $\rho$, and $\rho>175$~\um~is compatible with $\eta_T>95\%$ (shaded region). In practice, these two constraints are technically challenging to achieve consistently. 

The blue curves in Fig.~\ref{fig1:prms_geometry}(b)-(c) show the reflected fraction 
\begin{equation}
\label{eq:R_def}
     \eta_R = \left|\frac{\langle \vec E_\text{out}|\vec E_\text{out}\rangle }{\langle \vec E_\text{in}|\vec E_\text{in} \rangle}\right|^{2} = \left|\langle\vec E_\text{drive}|\vec{E}_{\text{out}}\rangle\right|^{2}
\end{equation}
where $\vec E_\text{drive}$ is the ``driving'' field just inside the surface (see Fig.~\ref{fig1:prms_geometry}(a)) generated by Fresnel reflection of the incident mode field $\vec E_\text{in}$ from the index change at the surface (note $\langle \vec E_\text{in}|\vec E_\text{in} \rangle = 1$, so $\langle \vec E_\text{drive}| \vec E_\text{drive}\rangle < 1$), and $\vec E_\text{out}$ is the outbound mode. For the same two situations, we can approximate
\begin{equation}
\label{eq:R_vs_theta_rho}
\begin{aligned}\eta_R\left(\theta\right)\approx &  \eta_{R,\text{max}} e^{-\sigma^{2}\beta^{2}\tan^{2}\theta}\\
\eta_R\left(\rho\right) \approx &  \frac{\eta_{R,\text{max}}}{1+\frac{\sigma^{4}\beta^{2}}{4\rho^{2}}}
\end{aligned}
\end{equation}
to leading order, where $\eta_{R,\text{max}}$ is the maximum reflection (appendix \ref{app:R}). Both the blue and red curves (Eq.~\ref{eq:R_vs_theta_rho}) are monotonic, linking $\eta_T$ to $\eta_R$, though with just back-reflection it is not possible to tell how much of the changes in $\eta_R$ come from $\theta$ versus $\rho$. As discussed below, we can numerically separate the effects of $\theta$ and $\rho$ using the measured value of $\rho$ from a profilometer to constrain $\theta$ and $\eta_T$. As shown below, however, it turns out that $\eta_R$ alone rapidly places tight bounds on the maximum achievable value of $\eta_T$ in the high-$\eta_T$ limit.

\section{Surface Characterization During Fabrication}
\label{sec:surf_char}
\begin{figure*}[htb]
\includegraphics[width=12cm]{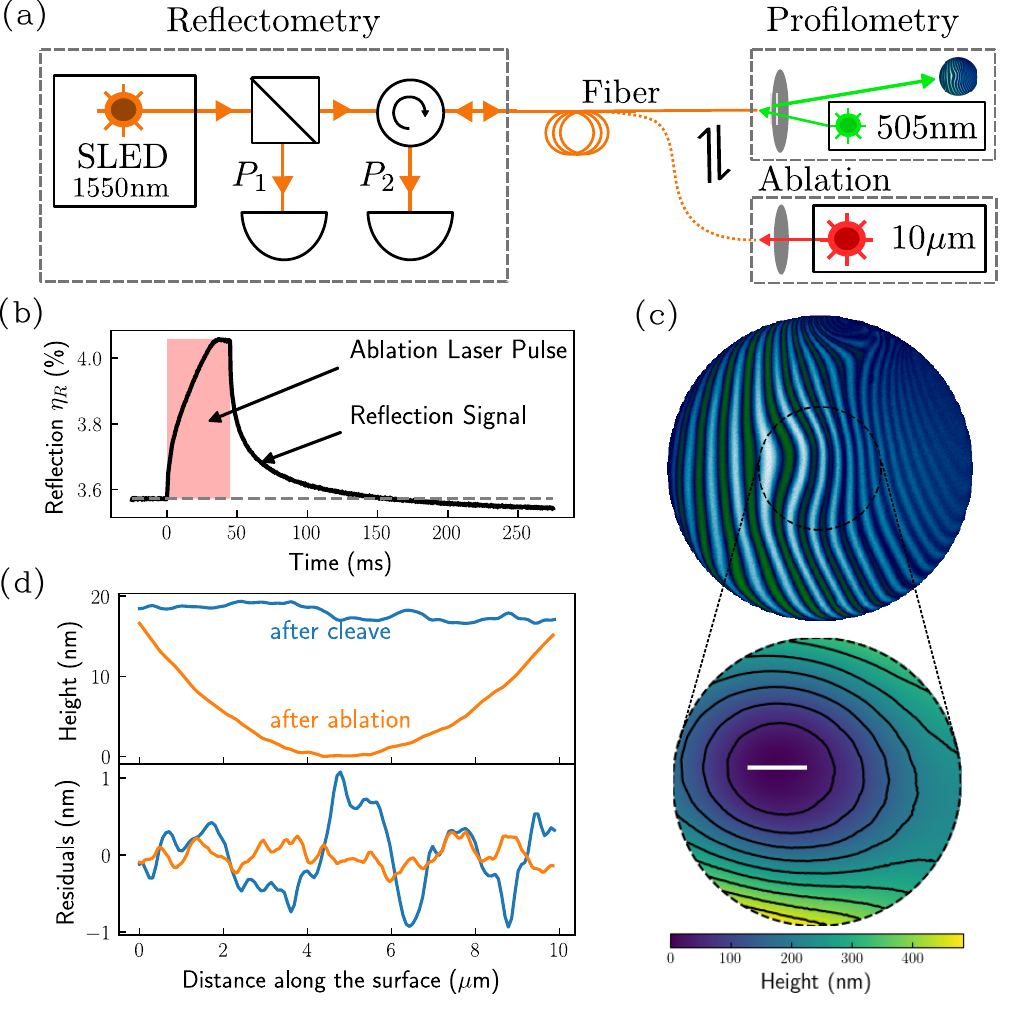}
\caption{
\label{fig2:ExpSetup}
\textbf{Experimental setup} 
(a) Apparatus for ablating the fiber tip and estimating the achievable coupling efficiency.
The reflectometry measurement uses light from a superluminescent light-emitting diode (SLED), which has a coherence length below $\sim$7 \um to eliminate standing-wave interference. The first photodiode (collecting power $P_1$) monitors power fluctuations, and the second (power $P_2$) measures light returning from the fiber tip. 
The end fiber of this \textit{reflectometer} is spliced into a fiber spool from which the other end is mounted on a translation stage to move between ablation and profilometry. 
(b) Measured reflection $\eta_R$ during ablation. The red fill
indicates when the CO$_2$ laser is on, consistent with a temporary increase (and decay) in refractive index commensurate with
the changing surface temperature. The gray dashed line is a guide set at the initial value.
(c) (top) Typical interferometric image, from a Mirau objective, of the fiber tip after ablation. 
(c) (bottom) Reconstructed profile within the 24~\um radius drawn in the top image. 
Dark lines are contour levels separated by 50~nm. 
The middle bar indicates the approximate 10~\um path of the AFM shown in (d). The apparent offset in the ablation minimum illustrates the difficulty interpreting profilometer images in the presence of systematic uncertainties in fiber tilt. (The back-reflection measurements circumvent this issue.)  
(d) AFM profiles near the center of the fiber tip after cleaving and ablation. Residuals (below) are estimated by subtracting a third-order polynomial fit. 
}
 
\end{figure*}

Figure~\ref{fig2:ExpSetup}(a) shows the combined optical reflection and profilometry systems used to characterize the substrate surface during fabrication. The profilometer is an imaging Mirau objective \cite{henri_mirau_1952} illuminated by a 505 nm LED source (50~nm bandwidth), which provides a contour height map of the fiber tip with fringes every 252.5 nm (inset image and (c,top)). Scanning the fiber position along the z-axis (parallel to the core) allows surface reconstruction \cite{wu_development_2021}, an example of which is shown in (c). 
% Now go with Figure~\ref{fig2:ExpSetup}(b) 
The CO$_2$ laser delivers 0.7~W for 45 ms with a mode field diameter of $\sim$100~\um. Figure \ref{fig2:ExpSetup}(b) shows the typical evolution of the reflection during ablation. The rise in $\eta_R$ during ablation is consistent with a temperature-dependent increase in refractive index, which will be the subject of future work. In all but a few cases, the post-ablation value falls below the post-cleave value (gray dashed line), consistent with our ablation generally decreasing the radius of curvature or introducing additional errors in $\theta$ (see Fig.~\ref{fig2:ExpSetup}(c)).

Figure~\ref{fig2:ExpSetup}(d) shows a ``typical'' atomic force microscope scan along the horizontal line after cleaving (blue) and again after ablation (orange). We subtract a third-order polynomial fit to recover residuals plotted below, which can be used to estimate roughness. Consistent with literature \cite{hunger2010a}, the ablation process decreases the post-cleave RMS roughness from 0.45~nm (blue) to 0.14~nm (orange).

In practice, the profilometer is well-suited to estimating the radius of curvature $\rho$, but cannot reliably estimate the tilt angle $\theta$ of the surface at the fiber core, due to image resolution, $\sim$\um~tolerances on the manufactured core location, the random nature of cleaves near the edge of the cladding, the misalignment of the Mirau's reference plane, and the residual $\sim$m bend radius of fibers from a spool (if a $\sim$mm-long fiber segment hangs freely, this adds $\sim$milliradian tilt in a random direction). 

To more reliably constrain the surface flatness, we perform a continuous reflectometry measurement shown in Fig.~\ref{fig2:ExpSetup}(a,b) during the cleave and ablation process. A superluminescent light-emitting diode (SLED) sends 10~mW of 1550~nm light (coherence length $\lesssim$ 7 \um~to eliminate standing-wave interference) through a 90:10 splitter, sending $P_1\sim$1~mW to a reference photodiode, and the rest through a circulator to the fiber tip. After component losses, the total power delivered to the fiber tip is 0.8~mW. 
The reflected power $P_2$ is collected through the circulator, and the ratio $P_2/P_1$ (insensitive to drifts in SLED power) is then proportional to the back-reflection coefficient $\eta_R$ defined in Eq.~\ref{eq:R_def}. To calibrate this relationship, we apply Bayesian analysis to the distribution of $P_2/P_1$ values from multiple cleaves, allowing us to precisely constrain the value of $P_2/P_1$ corresponding to $\eta_{R,\text{max}}$ (Appendix \ref{app:error_R}). This approach yields a 95\% confidence interval below 0.004\% for the $\eta_R$ values shown in Fig.~\ref{fig3:stats_tolerance}(a). Note this confidence interval extends only \textit{downward} from the plotted points, but is much smaller than the marker size. Measurement noise on $P_1$ and $P_2$ lead to an additional statistical fluctuation of $\pm 0.001\%$.

\section{Constraining the Mode Matching Parameter}
\label{sec:result}
\begin{figure*}[htb]
\includegraphics[width=17cm]{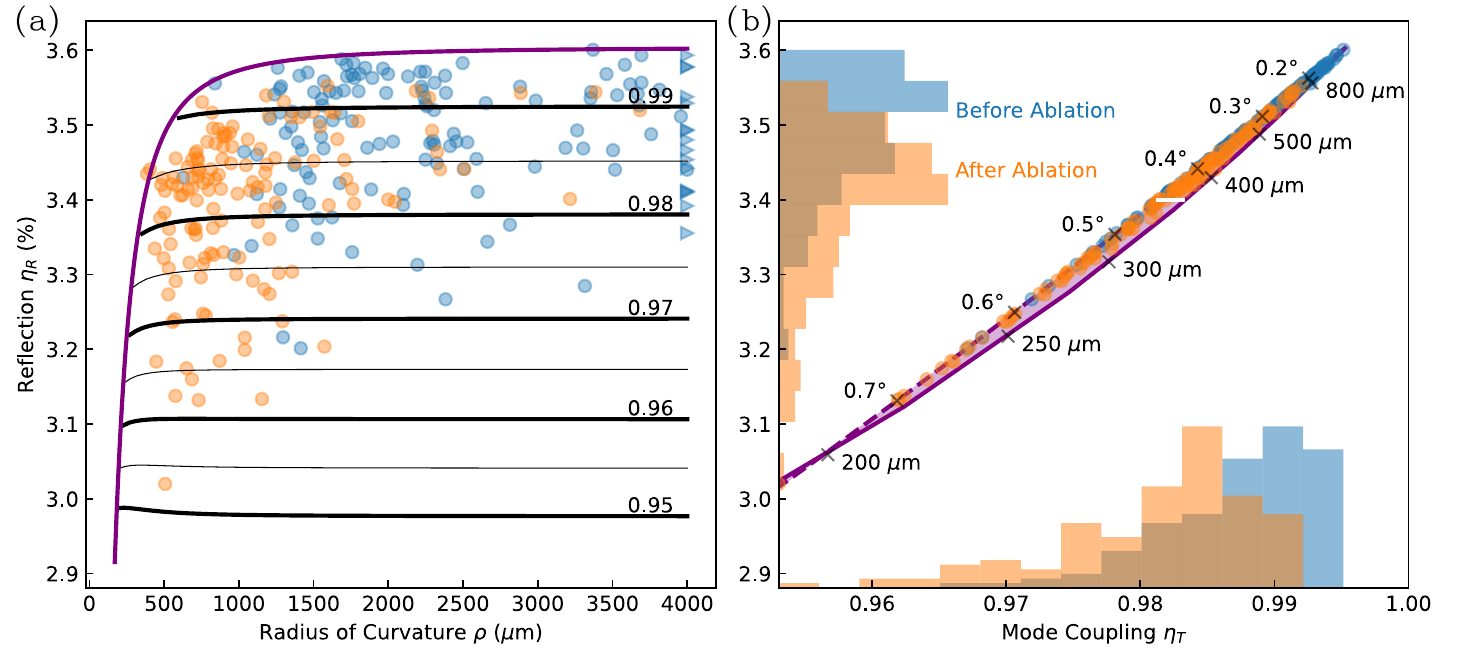}
\caption{
\label{fig3:stats_tolerance}
\textbf{Measured Fibers and Expected Coupling}
(a) Measured (smaller) radius of curvature ($\rho$) and back reflection ($\eta_R$) on 138 fibers before (blue markers) and after (orange markers) ablation. 
The violet curve is the theoretical limit for a perfect core alignment ($\theta=0$). The black lines are contours of constant mode coupling ($\eta_T$) with the indicated values (thin gray lines are half-values in between). The blue triangles indicate fibers with $\rho\geq4000\mu$m. 
(b) $\eta_R$ as a function of the mode coupling ($\eta_T$) for the same fibers with corresponding marginalized histograms on each axis. 
The solid violet line is the same limit line as in (a) with the marked values of $\rho$. The dashed purple line is the limit of infinite $\rho$ with the marked values of $\theta$. The filled purple region in between is the allowed value of $\eta_T$ and $\eta_R$. 
The white segment at $\eta_R=3.4\%$ illustrates an example bound on the achievable mode coupling, with $\eta_T=0.980\pm 0.001$.
}
\end{figure*}
After calibrating $\eta_R$, we selected 138 cleaved fibers for which $\eta_R$ was compatible with $\eta_T>0.96$.
Figure~\ref{fig3:stats_tolerance}(a) shows a scatter plot of $\eta_R$ (reflectometer) and $\rho$ (profilometer) for these fibers, after cleaving (blue) and after ablation (orange), showing the aforementioned trend toward less optimal surface profiles.
Since the surface profiles are generally slightly elliptical, we use the smaller radius of curvature for $\rho$, as this provides a lower bound on the achievable mode coupling $\eta_T$. The violet curve represents the theoretical limit in the absence of tilt ($\theta=0$), while the solid black lines represent contours of constant $\eta_T$ numerically calculated for each value of $\eta_R$ and $\rho$. 
For this we employ the full Bessel solutions of the fiber core (Appendix~\ref{app:R}). To estimate the ``maximum'' achievable $\eta_T$, the cavity mode was assumed to be a TEM$_{00}$ Gaussian mode centered on the core with a cavity axis perpendicular to the surface, and a width matched to that of the fiber core mode (Fig.~\ref{fig1:prms_geometry}(a)).
With this simple ablation protocol, all fibers ended up having surfaces compatible with $\eta_T>0.95$. The ability to pre-select substrates prior to coating is the first main result of this work.

Moreover, measuring the radius of curvature -- a comparatively cumbersome step -- is not strictly necessary in this parameter regime. Figure~\ref{fig3:stats_tolerance}(b) shows the same points versus $\eta_R$ and $\eta_T$ with histograms on each axis. Here the solid violet curve again represents the theoretical limit for $\theta=0$, while the dashed violet curve shows the limit for $\rho\rightarrow\infty$. The region bounded by these two curves is quite narrow, meaning a measurement of $\eta_R$ alone (i.e., a horizontal line cut) already strongly constrains $\eta_T$ in this parameter regime. For example, if we measure $\eta_R=3.4\%$, then we know the surface profile is compatible with $\eta_T=0.980\pm 0.001$ (white line). The ability to constrain the maximum achievable $\eta_T$ with a rapid reflectometry measurement is the second main result of this work.

\section{Conclusions}

We demonstrated a means to quantify surface quality during fiber mirror substrate fabrication, enabling pre-selection of substrates prior to coating. Interestingly, for those compatible with high coupling efficiency, a simple measurement of back-reflection provides rapid constraints without the need for profilometry.
While monitoring back-reflection, the process of cleaving and ablating takes only a few minutes, meaning we could in principle tighten the constraints on $\eta_R$, e.g., keeping only the $\sim$1\% of fibers compatible with $\eta_T>0.99$. 
The substrate yield could be further improved beyond this by optimizing the ablation pulse \cite{ruelle_optimized_2019}, employing a multi-shot approach \cite{ott_millimeter-long_2016}, or using an adaptive protocol \cite{gao_profile_2025}, which could open the possibility of fabricating flat and polished internal end facets to improve the lifetime of fiber-based quantum memories \cite{bustard_toward_2024}.

\section{Acknowledgments}
%\textcolor{red}{The authors thank Patrick Braganca and/or Brandon Ruffalo for assistance in fabricating nanomagnetic devices.} 
MCM acknowledges financial support from Fond de Recherche du Québec (FRQ-NT 318485). 
JCS acknowledges financial support from the Natural Sciences and Engineering Research Council of Canada (NSERC RGPIN 2018-05635 \& 2024-04620, ALLRP 578464-22), Canada Research Chairs (CRC 235060), NSERC Canadian foundation for Innovation (CFI 228130, 36423), Institut Transdisciplinaire d'Information Quantique (INTRIQ), and the Centre for the Physics of Materials (CPM) at McGill.
Special thanks to: Sean Cheng for invaluable help on the AFM.

\section{Data Availability}
Data from all figures and relevant scripts will be made freely available on the McGill Quantum Optics and Sensing Lab Dataverse (https://borealisdata.ca/dataverse/quantum\_optics\_and\_sensing).

\appendix

\section{Core light reflection coefficient}
\label{app:R}
Here we calculate the leading-order reflected power fraction $\eta_R$ (Eq.~\ref{eq:R_def}) for an arbitrary surface geometry at the fiber tip, then consider leading-order perturbations from flat.
Following common conventions, we write the incident wave (propagating along $\hat{z}$ with single polarization $\hat{x}$) as
\begin{equation}
\begin{aligned}
\vec{E}_{\text{in}}(x,y,z)&=f(x,y)e^{+i\beta z}\hat{x}\\
\end{aligned}
\end{equation}
where $f(x,y)$ is the core mode profile normalized so that $\iint dxdy \left|f\left(x,y\right)\right|^{2}=1$ and $\beta$ is the propagation constant of the fiber (i.e., the effective wavenumber along $\hat{z}$ including the index of refraction and confinement effects \cite{saleh_fundamentals_2019}). 
The reflected power from Eq.~\ref{eq:R_def} is found by calculating a field $\vec E_\text{drive}$ (that ``drives'' the returning light) from $\vec{E}_{\text{in}}$ by locally analyzing each point $(x,y)$ of the fiber tip's surface at height $z=S(x,y)$, such that the area $dxdy$ is small enough to view the surface as flat but tilted to leading order; we then apply Fresnel's reflection laws to calculate how the incident field $\vec{E}_{\text{in}}$ is reflected at that point. Treating $s$ and $p$ polarization components separately,
\begin{equation}
\vec{E}_{\text{drive}}=r_{s}\left(\vec{E}_\text{in}\cdot\hat{s}\right)\hat{s}+r_{p}\left(\vec{E}_\text{in}\cdot\hat{p}\right)\hat{p}
\end{equation}
with in-plane unit vectors $\hat{s}=\frac{\left(-S_{y},S_{x},0\right)}{\sqrt{S_{x}^{2}+S_{y}^{2}}}$ and $\hat{p}=\frac{\left(S_{x},S_{y},0\right)}{\sqrt{S_{x}^{2}+S_{y}^{2}}}$, and surface derivatives $S_{x,y}\equiv\partial_{x,y}S$. The corresponding reflection coefficients are then
\begin{equation}\label{eq:rs-rp}
\begin{aligned}
r_{s}=&\frac{n_{\text{f}}\cos\theta_{i}-n_{\text{a}}\cos\theta_{t}}{n_{\text{f}}\cos\theta_{i}+n_{\text{a}}\cos\theta_{t}},\\r_{p}=&\frac{n_{\text{f}}\cos\theta_{t}-n_{\text{a}}\cos\theta_{i}}{n_{\text{f}}\cos\theta_{t}+n_{\text{a}}\cos\theta_{i}},\\\cos\theta_{i}=&\frac{1}{\sqrt{1+S_{x}^{2}+S_{y}^{2}}}\\\cos\theta_{t}=&\sqrt{1-\left(n_{\text{f}}/n_{\text{a}}\right)^{2}\left(1-\cos^{2}\theta_{i}\right)},
\end{aligned}
\end{equation}
where $n_\text{f}(x,y)$ is the refractive index for the fiber at location ($x,y$), and $n_\text{a}$ is the index of the surrounding medium (i.e., air). 

The returning power is the sum of the power coupling into each $\hat{x},\hat{y}$ polarization mode, such that the \textit{total} reflected power fraction $\eta_R=\eta_{x}+\eta_{y}$, with
\begin{equation}
\begin{aligned}
\eta_{x}=&\left|\left\langle \vec{E}_{\text{drive}}|f(x,y)e^{-i\beta z}\hat{x}\right\rangle \right|^{2}\\=&\left|\iint dxdyf\left(x,y\right)^{2}e^{-2i\beta S\left(x,y\right)}\frac{r_{p}S_{x}^{2}+r_{s}S_{y}^{2}}{S_{x}^{2}+S_{y}^{2}}\right|^{2}\\\eta_{y}=&\left|\left\langle \vec{E}_{\text{drive}}|f(x,y)e^{-i\beta z}\hat{y}\right\rangle \right|^{2}\\=&\left|\iint dxdyf\left(x,y\right)^{2}e^{-2i\beta S\left(x,y\right)}\frac{\left(r_{p}-r_{s}\right)S_{x}S_{y}}{S_{x}^{2}+S_{y}^{2}}\right|^{2}.
\end{aligned}
\end{equation}
For surfaces with small gradients ($\left|S_{x}\right|,\left|S_{y}\right|\ll1$) we can simplify by expanding $r_{s},r_{p}$ (Eqs.~\ref{eq:rs-rp}) to first order in the quantity $S^{2}_{x}+S^{2}_{y}$, leading to 
\begin{equation}
    \begin{aligned}
        r_{s}\approx&r_{0}\left(1+\left(S^{2}_{x}+S^{2}_{y}\right)\frac{n_{f}}{n_{a}}\right)\\r_{p}\approx&r_{0}\left(1-\left(S^{2}_{x}+S^{2}_{y}\right)\frac{n_{f}}{n_{a}}\right),
    \end{aligned}
\end{equation}
with $r_{0}\equiv \frac{n_{f}-n_{a}}{n_{a}+n_{f}}$. In this case, to zero'th order in $S_{x}, S_{y}$, we have: 
\begin{equation}
\begin{aligned}
\eta_R\approx&\left|\iint dxdyf\left(x,y\right)^{2}e^{-2i\beta S\left(x,y\right)}\frac{n_{\text{f}}\left(x,y\right)-n_{\text{a}}}{n_{\text{f}}\left(x,y\right)+n_{\text{a}}}\right|^{2}
\end{aligned}
\end{equation}
which is consistent with the thin film approximation \cite{bond_higher_2011, saleh_fundamentals_2019}. 

To obtain a closed-form expressions for $\eta_R$, we assume that the reflection factor, $\frac{n_{\text{f}}\left(x,y\right)-n_{\text{a}}}{n_{\text{f}}\left(x,y\right)+n_{\text{a}}}$, have a fixed averaged value, $\sqrt{\eta_{R,\text{max}}}$ that corresponds to the maximum observable reflection:
\begin{equation}
    \label{eq:averaged_reflection}
    \eta_{R,\text{max}}\equiv \left|\iint dxdyf\left(x,y\right)^{2}\frac{n_{\text{f}}\left(x,y\right)-n_{\text{a}}}{n_{\text{f}}\left(x,y\right)+n_{\text{a}}}\right|^{2}
\end{equation}

For our step-index telecom fiber (SMF-28 Ultra) at 1550 nm, using $n_{\text{f}}=\{1.4700, 1.4648 \}$, a core radius of 4.1\um and the Bessel solutions for the single core mode \cite{saleh_fundamentals_2019}, $\eta_{R,\text{max}}$ converges to 3.604\%. This reflection gives an averaged index of 1.4687 --- consistent with our manufacturer specified effective group index of refraction.

We can also get analytical solutions for leading-order perturbations from a flat surface, approximate the surface near the core by $S\left(x,y\right)=\frac{x^{2}+y^{2}}{2\rho}+x\tan\theta$, and approximating the core mode with a Gaussian $f\left(x,y\right)=\frac{1}{\sigma}\sqrt{\frac{2}{\pi}}e^{^{-(x^{2}+y^{2})/\sigma^{2}}}$, yielding
\begin{equation}
\begin{aligned}
\label{eq:eta_R_approx}
\eta_R\approx&\frac{\eta_{R,\text{max}}}{1+\frac{\sigma^{4}\beta^{2}}{4\rho^{2}}}e^{-\frac{\sigma^{2}\beta^{2}\tan^{2}\theta}{1+\frac{\sigma^{4}\beta^{2}}{4\rho^{2}}}}
\end{aligned}
\end{equation}

\section{Bayesian Estimation of Reflection Coefficient}
\label{app:error_R}
\begin{figure}
\includegraphics[scale=0.55]{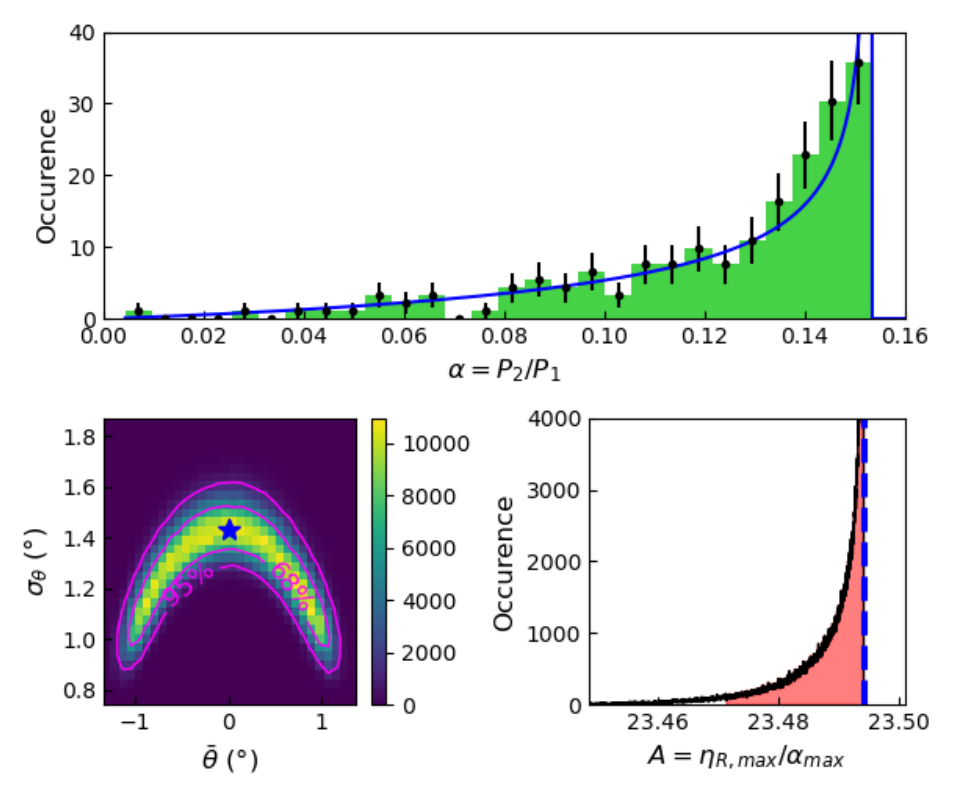}
\caption{
\label{fig:cleave_stats}
\textbf{Bayesian calibration of reflectometer}
(a) Histogram of 137 measured post-cleave values of $\alpha\equiv P_2/P_1$.
The solid curve is the likelihood modeled by Eq.~\ref{eq:Likelihood} (normalized) with the parameters taken as the most likely values $(\amax=0.1534, \bar\theta=0^\circ,\sigma_\theta=1.43^\circ)$ sampled in (b) and (c). Uncertainty bars show the square root of counts for reference (not used in the analysis). 
(b) 
2D histogram of the Monte-Carlo-sampled “nuisance parameters” $\bar{\theta}$ and $\sigma_{\theta}$. Contour lines enclose the 68\% and 95\% confidence intervals. The marker indicates the values used in (a).
(c)
Histogram of the calibration parameter $A=\eta_{R,\text{max}}/\alpha_{\text{max}}$, with $\alpha_{\text{max}}$ sampled together with $\bar\theta$ and $\sigma_\theta$ as shown in (b). 
The red shaded region indicates the 95\% confidence interval from the maximum value (vertical dashed line at $A=23.494$).
}
\end{figure}
The measured quantity $\alpha\equiv P_1/P_2$ is directly related to the reflection coefficient through $\eta_R=A\alpha$, where $A$ is a calibration constant to be determined. One can in principle calibrate $A$ by carefully characterizing the losses of all components along the optical paths, but in practice this does not sufficiently constrain $A$ to make meaningful conclusions about the data in Fig.~\ref{fig3:stats_tolerance}, primarily due to the irreproducibility of fiber connections, which place a 10\% systematic uncertainty on $A$. Here we instead apply Bayesian inference to estimate $A$ (and hence $\eta_R$) using just the observed post-cleave distribution of $\alpha$.

Figure~\ref{fig:cleave_stats}(a) shows a histogram of measured $\alpha$ values from the 137 cleaves (before ablation) discussed in this manuscript. The distribution suggests a cutoff at an unknown value $\amax$, determined by the maximum possible back-reflection $\eta_{R,\text{max}}$ (Appendix \ref{app:R}), such that $A=\frac{\eta_{R,\text{max}}}{\amax}$. For simplicity (and because it captures the observed distribution), we assume the dominant cause of fluctuations in $\alpha$ is from the cleave angle $\theta$ — the conservative pre-ablation radius of curvature shown in Fig.~\ref{fig3:stats_tolerance}(a) is generally well above 1000~\um, where it does not strongly affect $\eta_R$ — and we assume a Gaussian distribution
\begin{equation}
    \mathcal{P}(\theta)=\frac{1}{\sigma_{\theta}\sqrt{2\pi}}e^{-(\theta-\bar{\theta})^2/2\sigma^{2}_{\theta}}
\end{equation}
with mean $\bar\theta$ and standard deviation $\sigma_\theta$. 
Mapping this onto Eq.~\ref{eq:R_vs_theta_rho} (dividing \textit{both} sides by $A$ so that $\alpha_i=\alpha_{\text{max}}e^{-\sigma^{2}\beta^{2}\tan^{2}\theta}$) and noting that both solutions of $\theta$, $\theta_{\pm}=\pm\arctan\left(\frac{1}{\sigma\beta}\sqrt{\ln\left(\frac{\alpha_{\text{max}}}{\alpha_{i}}\right)}\right)$, contribute to $\alpha_i$, we arrive at a ``model'' probability distribution function
\begin{align}
\label{eq:Likelihood}
\mathcal{P}\left(\alpha_{i}|\alpha_{\text{max}},\bar{\theta},\sigma_{\theta}\right)=&\mathcal{P}\left(\theta_{+}\right)\left|\frac{d\theta_{+}}{d\alpha_{i}}\right|+\mathcal{P}\left(\theta_{-}\right)\left|\frac{d\theta_{-}}{d\alpha_{i}}\right|\\=&\frac{e^{-(\theta_{+}-\bar{\theta})^2/2\sigma^{2}_{\theta}}+e^{-(\theta_{-}-\bar{\theta})^2/2\sigma^{2}_{\theta}}}{2\sigma\beta\alpha_{i}\sigma_{\theta}\sqrt{2\pi\ln\left(\frac{\alpha_{\text{max}}}{\alpha_{i}}\right)}},
\end{align}
describing the likelihood of measuring $\alpha_i$ given the calibration parameter $\alpha_\text{max}$ and ``nuisance'' parameters $\bar \theta$ and $\sigma_\theta$. According to Bayes' relationship, the probability density function of the three unknown variables given a set of $N=137$ independent measurements $\{\alpha\}$ is then \cite{sivia_data_2006}
\begin{align}\label{eq:P3D-given-measurements}
&\mathcal{P}(\alpha_{\text{max}},\bar{\theta},\sigma_{\theta}|\{\alpha\})\\
&\propto\mathcal{P}(\{\alpha\}|\alpha_{\text{max}},\bar{\theta},\sigma_{\theta})\mathcal{P}(\alpha_{\text{max}},\bar{\theta},\sigma_{\theta})\\
&\propto\prod^{N}_{i=1}\mathcal{P}(\alpha_{i}|\alpha_{\text{max}},\bar{\theta},\sigma_{\theta})\mathcal{P}(\alpha_{\text{max}})\mathcal{P}(\bar{\theta})\mathcal{P}(\sigma_{\theta}),
\end{align}
where we assume independent uniform prior probability densities $\mathcal P(\amax)$ spanning $0<\amax<1$, $\mathcal P(\bar\theta)$ spanning $-1<\bar\theta<1$ radian, and $\mathcal P(\sigma_\theta)$ spanning $0<\sigma_\theta<1$ radian. These ranges conservatively span much more parameter space than we expect from the aforementioned $\pm 10\%$ accurate calibration based on component losses.
%Since our initial estimate of $A$ (based on calibrating component losses) has a 10\% systematic uncertainty, we conservatively assume a uniform prior for $\amax$ spanning $\pm 0.5$ around the nominal value $a=0.5$. The priors for $(\bar\theta,\sigma_\theta)$ are also flat, spanning $-1$ to $+1$\,rad for $\bar\theta$ and $0$ to $+1$\,rad for $\sigma_\theta$. 

% MCMC
We sample $(\amax,\bar\theta,\sigma_\theta)$ from Eq.~\ref{eq:P3D-given-measurements} by using an affine-invariant Markov Chain Monte Carlo (MCMC) ensemble sampler \cite{emcee2013}, which naturally provides marginalized distributions for all the variables. The results from 32 ``walkers'', 500 initial ``burn-in" steps, and 5000 sampled steps — a total of 160k samples — are shown in Fig.~\ref{fig:cleave_stats}(b,c). In (b), the two-dimensional histogram of the ``nuisance'' cleaver parameters is shown together with confidence intervals. These variables explore a wide range of values, but the 95\% confidence interval still tightly constrains the calibration parameter shown in (c) to within 0.1\% below the most likely value. This corresponds to an interval spanning less than 0.004\% below the values of $\eta_R$ plotted in Fig.~\ref{fig3:stats_tolerance}.
%
%\bibliography{Main, fiber_fab_michael} % Produces the bibliography via BibTeX.
\providecommand{\noopsort}[1]{}\providecommand{\singleletter}[1]{#1}%
%
 % Switch to this for submission.

\end{document}